\documentclass[useAMS,usenatbib,onecolumn]{mn2e}
\usepackage{epsfig}
\usepackage{amsmath}
\usepackage{amssymb}
\newcommand{\abs}[1]{\left|#1\right|}           

\title[Instabilities of captured  shocks]
{Instabilities of captured  shocks  in the envelopes of massive stars}
\author[M. Grott et al.]
{M. Grott$^{1}$\thanks{E-mail: mgrott@uni-sw.gwdg.de (MG)}, 
W.Glatzel$^{1}$ and 
S. Chernigovski$^{2}$ \\
$^{1}$Universit\"ats-Sternwarte G\"ottingen, Geismarlandstra{\ss}e 11, 
D-37083 G\"ottingen, Germany\\
$^{2}$Institut f\"ur Analysis und Numerik, Universit\"at Magdeburg, 
Universit\"atsplatz 2, D-39106 Magdeburg, Germany}
\begin{document}


\pagerange{\pageref{firstpage}--\pageref{lastpage}} \pubyear{2002}

\maketitle

\label{firstpage}
 
\begin{abstract}
The evolution of strange mode instabilities into the 
non linear regime has been followed by numerical simulation for an envelope model
of a massive star having solar chemical composition, $M=50M_\odot$, 
$T_{\rm{eff}}=10^4K$ and $L=1.17\cdot 10^6 L_\odot$. Contrary to previously
studied models, for these parameters shocks are captured in the H-ionisation zone 
and perform rapid oscillations within the latter.  
A linear stability analysis is performed to verify that this behaviour is physical.
The origin of an instability discovered in this way is identified by 
construction of an analytical model. As a result, the stratification turns out
to be essential for instability. The difference to common stratification 
instabilities, e.g., convective instabilities, is discussed.
\end{abstract}


\section{Introduction}

Massive stars are known to suffer from strange mode instabilities with growth 
rates in the dynamical range \citep{KFG93,GK93}. 
The boundary of the domain in the Hertzsprung-Russel diagram (HRD) 
where all stellar models are unstable - irrespective of their 
metallicity -, coincides with the observed Humphreys-Davidson (HD) 
limit \citep{HD79}. Moreover, the range of unstable models covers the 
stellar parameters for which the LBV (luminous blue variable) 
phenomenon is observed.  

The high growth rates of the instabilities indicate a connection to 
the observed mass loss of the corresponding objects. To verify this suggestion,
simulations of their evolution into the non linear regime have been performed.
In fact, for selected models \citet{GKCF99} found 
the velocity amplitude to exceed
the escape velocity (see, however, \citet{DG00}).

In this paper we report on a stellar model, which  in the HRD is located 
well above the HD-limit, however, at lower effective temperature than the model 
studied by  \citet{GKCF99}. 
As expected, this model turns out to be linearly unstable with 
dynamical growth rates.  When
following the non linear evolution of the instabilities, shocks form
in the non linear regime.
The latter is customary in pulsating stellar envelopes
(see, e.g., \citet{C66}). Contrary to the ``hotter'' model studied by
\citet{GKCF99}, however, these shocks are captured by the H-ionisation zone
after a few pulsation periods. The captured shock starts to oscillate rapidly with
periods of the order of the sound travel time across the H-ionisation zone, 
while its mean position changes on the dynamical timescale of the primary,
strange mode instability. This phenomenon is described in detail in section
3.1. Assumptions and methods on which the calculations are based are given
in section 2. 
We emphasise, that in this publication, we concentrate on the oscillations
of the captured shock. The phenomenon of shock capture by H-ionisation itself
is not investigated here and will be studied in a separate paper.

Apart from a detailed description of the shock oscillations found by 
numerical simulation the aim of the present paper consists of identifying
their origin. This will be achieved by a linear stability analysis in section
3.2. It excludes a numerical origin and attributes the oscillations to a 
secondary high frequencey instability in the shock zone. To identify the
physical origin of the instability an analytical model is constructed in
section 4. Our conclusions follow.

\section{BASIC ASSUMPTIONS AND METHODS}
\subsection{Construction of initial model}

We investigate a stellar model having the mass $M=50M_\odot$, 
chemical composition $X=0.7$, $Y=0.28$, $Z=0.02$, effective temperature
$T_{\rm{eff}}=10^4K$ and luminosity $L=1.17\cdot 10^6 L_\odot$. 
These parameters have been chosen to ensure instability of the model.
In the Hertzsprung-Russell diagram (HRD)
it lies within the instability region identified by \citet{KFG93}
(c.f. their figure 2). 
As only the envelope is affected by the instability,
the model was constructed by standard 
envelope integration using the parameters given above. 
The stellar core and nuclear energy generation are disregarded.
Convection is treated in the standard mixing-length theory approach with 1.5 pressure
scaleheights for the mixing length. The onset of convection was determined
by the Schwarzschild 
criterion. For the opacities, the latest versions of the 
OPAL tables \citep{IRW92,RI92} have been used.  

\subsection{Linear stability analysis}
\label{LNA}

Having constructed a hydrostatic envelope model its stability with 
respect to infinitesimal, spherical perturbations is tested.
The relevant equations corresponding to mass, energy and
momentum conservation and the diffusion equation for energy
transport are given in \citet{BK62} (hereafter BKA):
\begin{eqnarray}
  &&\zeta' = C_4(3\zeta +C_5p-C_6t)  \label{perturbation1}\\
  && l'     = (i\sigma)C_1(-p +C_2t) 
  \label{perturbation2}\\
  &&p'         = -(4+C_3\sigma^2)\zeta -p \label{perturbation3}\\
  &&t'         = C_7(-4\zeta+C_{13}l+C_8p-C_9t) 
  \label{perturbation4}
\end{eqnarray}
$\zeta$, $l$, $p$ and $t$ are the relative perturbations of radius, luminosity,
pressure and temperature, respectively, and dashes denote derivatives with respect to 
$\ln p_0$. $\sigma$ is the eigenfrequency normalized to the inverse of the
global free fall time $\tau_{\rm{ff}}=\sqrt{R^3/3GM}$. 
The coefficients $C_i$ are determined by the background model where
$C_{13}$ denotes the ratio of total and radiative luminosity. The other 
cofficients are defined in BKA.
For the general theory of linear non-adiabatic stability, we refer 
the reader to \citet{C80} and \citet{U89}. 

The coupling between 
pulsation and convection is treated in the standard frozen in 
approximation, i.e., 
the Lagrangian perturbation of the convective flux is assumed to vanish. This is 
justified since the convective flux never exceeds 10\% of the total energy flux.
Moreover, the convective timescale is much longer than the dynamical timescale of
the pulsations considered.
The solution of the perturbation problem has been determined using the 
Riccati method \citep{GG90}. As a result of the linear 
non adiabatic (LNA) stability
analysis we obtain periods and growth or damping rates 
of various modes together with the 
associated eigenfunctions.

\subsection{Non-linear evolution}

Having identified an instability by the LNA analysis its growth
is  followed into the non-linear regime. 
Assuming spherical symmetry, we adopt a Lagrangian description and choose
as independent
variables the time t and the mass $m$ inside a sphere of radius $r$. The 
evolution of an instability is then governed by  
mass conservation,
\begin{eqnarray}
  \frac{\partial r^3}{\partial m} -\frac{3}{4\pi\rho}&=&0
  \label{nl1}
\end{eqnarray}
momentum conservation,
\begin{eqnarray}
  \frac{\partial^2 r}{\partial t^2} +4\pi r^2 
  \frac{\partial P}{\partial m} +\frac{Gm}{r^2}&=&0
  \label{nl2}
\end{eqnarray}
energy conservation,
\begin{eqnarray}
  \frac{\partial L}{\partial m}- \frac{P}{\rho^2}
  \frac{\partial \rho}{\partial t}+\frac{\partial E}{\partial t} &=&0
  \label{nl3}
\end{eqnarray}
and the diffusion equation for energy transport,
\begin{eqnarray}
  \frac{\partial T}{\partial m}-\frac{3\kappa(L-L_{konv})}{64\pi^2acr^4T^3} &=&0 
  \label{nl4}
\end{eqnarray}
where $\rho$, $p$, $T$, $L$, and $E$ denote density, pressure, temperature, 
luminosity and specific internal energy, respectively. $a$ is the radiation
constant, $c$ the speed of light and $G$ the gravitational constant. 
For consistency, the equation of state 
$p(\rho,T)$ and the opacity $\kappa$ are identical with those used for 
the construction of the initial model. 
In accordance with the LNA stability analysis, convection is treated in the
frozen in approximation, i.e., $L_{konv}$ is taken to be constant during the
non-linear evolution and equal to the initial value. 
For the treatment of shocks artificial viscosity is introduced
by substituting $P=P+Q$ with ($v$ is the velocity)
\begin{eqnarray}
 Q= \begin{cases} 
    C_0\rho ({\rm{div }} \ v)^2 \qquad &{\rm{div}}\ v<0 \\
    0 \ &{\rm{div}}\ v\ge  0 
  \end{cases}
\end{eqnarray}
and $C_0>0$ 
{ 
(von Neumann - Richtmyer form of artificial viscosity).

For some difference schemes including the Fraley scheme, which the present 
method is based on,
this form of the artificial viscosity can give rise to undesired, unphysical
oscillations (see, e.g., \cite{B90}). 
To avoid these, artificial tensor viscosity is usually used
(\cite{TW79}). In order to be sure that the oscillations observed 
are not caused by the form of the artificial viscosity, we have run
tests both with volume and tensor viscosity. As a result, shock oscialltions
are found independently for any form of the artificial viscosity. As the
von Neumann - Richtmeyr viscosity allows for a straightforward formulation
of the boundary conditions discussed below, we have for 
convenience chosen to work with it.
}

The inert hydrostatic core provides boundary conditions at the 
bottom of the envelope by  prescribing its time independent radius and 
luminosity there.
As the outer boundary of the model does not correspond to the physical boundary
of the star, boundary conditions are ambigous there. 
We require the gradient of heat sources to vanish there:
\begin{eqnarray}
  {\rm{grad}}({\rm{div}} F)=0 
\end{eqnarray}
This boundary condition is chosen to ensure that outgoing shocks pass through the
boundary without reflection. 
{
The numerical code relies on 
a Lagrangian, with respect to time implicit, fully conservative 
difference scheme  proposed by   
\cite{FR68} and \cite{SP69}
}
Concerning tests of the code, we adopted the same criteria as \citet{GKCF99}. 

\section{NUMERICAL RESULTS}
\subsection{The evolution of the stellar model}
\label{evolution}
 
\begin{figure*}
  \epsfig{file=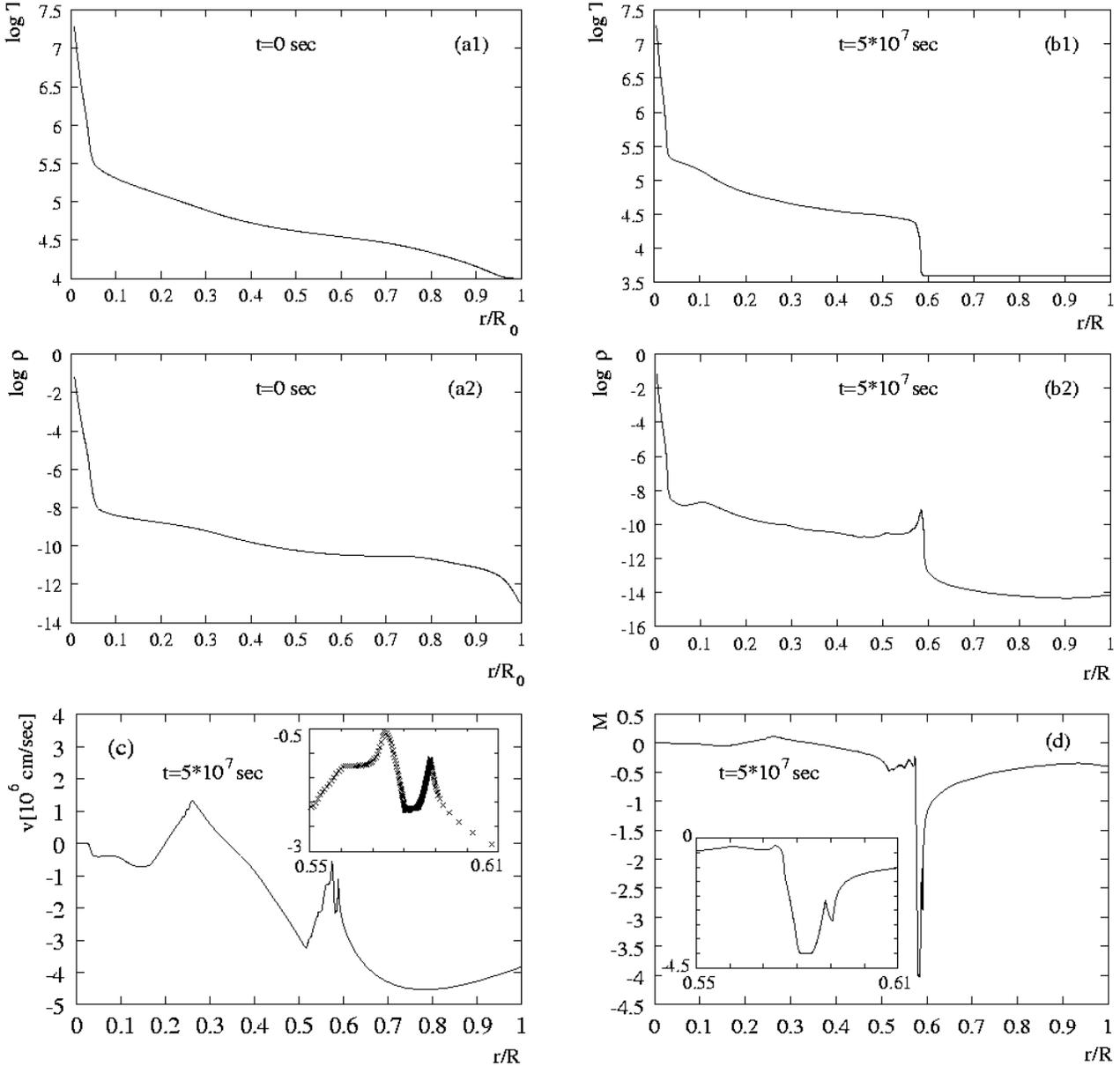,width=16.8cm}
  \caption{Temperature $T$ and density $\rho$ as a function of relative radius
    of the initial model (a1-a2), and the model at $5\cdot 10^7$ sec (b1-b2).
    {{Velocity $v$ and Mach number $M$ are shown as a function of relative
        radius for the model at $5\cdot 10^7$ sec in (c) and (d), 
        respectively}}}
  \label{start}
\end{figure*} 
  
Density and temperature of the initial model as a function of relative radius,
are shown in figures \ref{start}.a1-\ref{start}.a2. The stratification exhibits a  
pronounced core-envelope structure, which is typical for stellar models in this
domain of the HRD. More than $96$ per cent of the
mass is concentrated in the core, which extends to less than 5 per 
cent of the total radius. It remains in hydrostatic equilibrium and is
not affected by the instability.
 
The initial model has been tested for stability and been found unstable
on dynamical timescales.
This instability will be referred to as primary instability 
hereafter.  
With respect to its physical origin, it is a strange mode instability, which
has been identified in a variety of stars including
Wolf-Rayet-stars, HdC-stars and massive stars (like the present model).
Strange modes appear as mode coupling phenomena with associated instabilities
whenever radiation pressure is dominant. The latter is true for a large 
fraction of the radius in the present model.   
The linear stability analysis of the initial model reveals several unstable modes.
Eigenfrequencies of the most unstable ones, i.e., their real ($\sigma_r$) and
imaginary  parts ($\sigma_i$), are presented in table \ref{primary}. 
\begin{table}
  \caption{Unstable modes of the initial model.}
  \label{primary}
  \begin{tabular}{crrrrrrr}
    \hline
    $\sigma_r$ & 0.53 & 1.22 & 1.66 & 2.12 & 2.26 & 3.34 & 3.86 \\
    $\sigma_i$ & -0.06& -0.18& -0.13& -0.20& -0.12& -0.04& -0.04 \\
   \hline
  \end{tabular}

  \medskip
   $\sigma_r$  denotes the real part and  $\sigma_i$ the imaginary part of the 
   eigenfrequency $\sigma$ normalized by the global free fall time.
\end{table}

The evolution of the linear instabilities was followed into the non-linear regime
by numerical simulation
using the hydrostatic model as initial condition. No additional initial perturbation
of the hydrostatic model was added. Rather the code was required to pick the 
correct unstable modes from numerical noise. By comparing growth rates and periods
obtained in the simulation with the results of the LNA analysis, the linear regime
 of the evolution was used as a  test for the quality of the simulation.

\begin{figure*}
  \epsfig{file=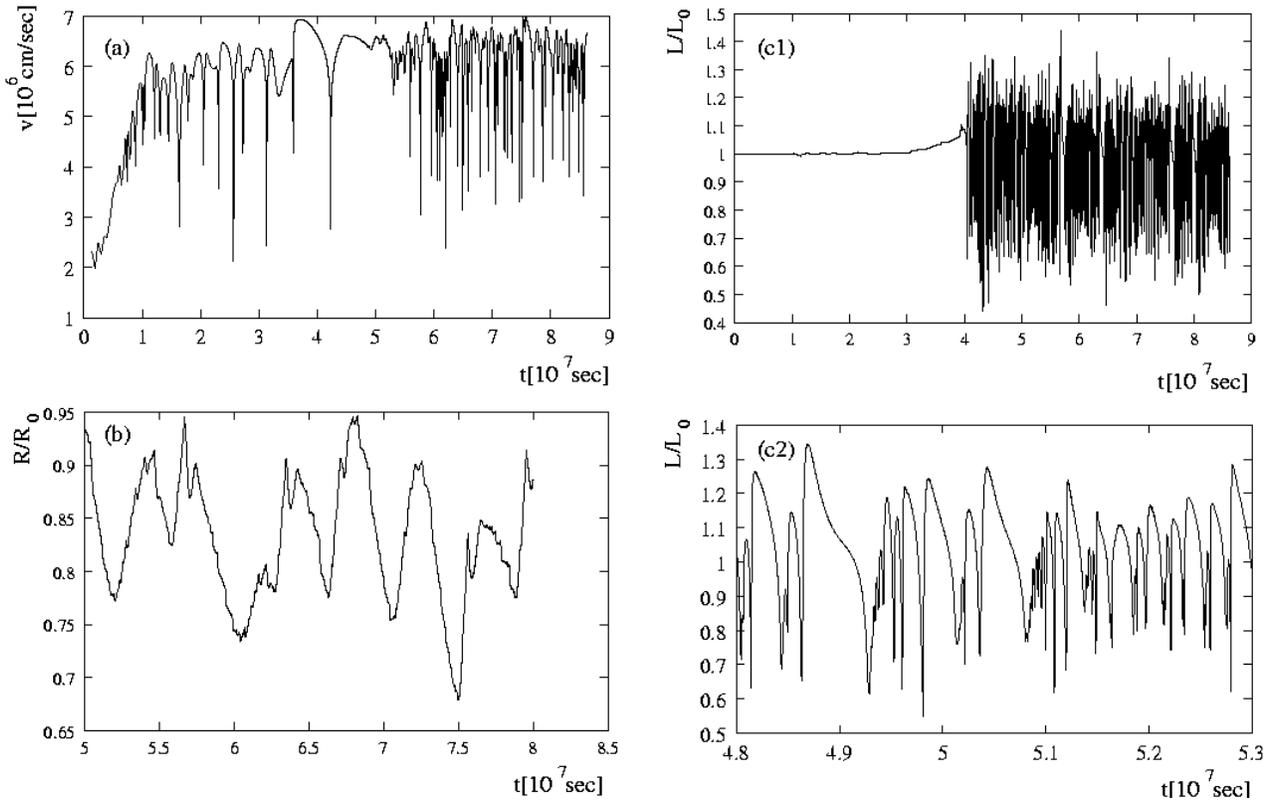,width=16.8cm}
  \caption{The velocity $v$  at the boundary (a), the relative 
    position of the shock front (b) and the relative luminosity $L/L_0$ at 
    the boundary on two different scales (c1-c2) as a function of time $t$.}
  \label{Time}
\end{figure*}

In the non-linear regime sound waves
travelling outwards form shocks and initially 
inflate the envelope to $2.5$ initial radii.
Thus velocity amplitudes of $10^7$[cm/sec] are reached.
One of the subsequent shocks is captured 
around the H-ionization zone at relative radius $r/R=0.58$ and 
$3.6< \log T < 4.7$. 
{
The mechanism responsible for the shock capturing will not be studied in  
this publication. Rather, we will investigate the oscillations
on the shock front and show that they are of physical origin.    
A snapshot at time $t=5\cdot 10^7$sec
of the situation containing the captured shock
is shown in Figures 
\ref{start}.b1 and \ref{start}.b2. 
Figure \ref{start}.c shows the velocity as a function of relative radius 
at this instant. Sound waves are generated in the
region around $r/R\approx 0.1$ and travel outwards, growing 
in amplitude and steepening. In the snapshot one such wave is located at
$r/R\approx 0.25$. The captured shock front is located at $r/R\approx 0.58$
and the outer envelope is collapsing onto it. The small panel in Figure 
\ref{start}.c shows the details of the region containing the captured shock, 
indicating the grid resolution by ($\times$). Within the Lagrangian 
description, $\sim 150$ of the 512 gridpoints used are concentrated in the 
shock zone. 
Figure \ref{start}.d shows the Mach number $M=v/v_s$ as a function of 
relative radius for the snapshot ($v_s$ is the local sound speed). 
The Mach number changes by 3.5 across the shock front around
$r/R\approx 0.58$. 

After the formation of the captured shock its position varies
} 
only weakly by $\approx 0.2$ relative radii on the timescale of the primary
instability (see figure \ref{Time}.b).  
Superimposed on this variation is a much faster oscillation, whose timescale
is related to the sound travel time across the shock ($\approx 10^5$ sec). 
It is even more
pronounced in the run of the luminosity (figure \ref{Time}.c2). The onset of the
fast oscillation with the capturing of the shock by the H-ionization-zone
is illustrated in figures \ref{Time}.a and \ref{Time}.c1, where the velocity $v$ and
relative luminosity $L/L_0$ at the outer boundary are shown as a function of time.
Up to $\approx 4\cdot 10^7$ sec the velocity varies on the timescale of the 
primary instability and the luminosity remains approximately constant due to 
the low heat capacity of the envelope of the star. 
After $\approx 4\cdot 10^7$ sec, when the shock has been captured by the 
H-ionization-zone, luminosity and velocity vary on the shorter timescale of the 
secondary shock oscillation. The luminosity perturbation has its
origin in the shock. Due to the low heat capacity the luminosity perturbation
remains spatially constant above the shock. 

In principle, the high-frequency secondary oscillations of the shock could be 
caused numerically. However, the results are largely independent of the numerical
treatment and parameters, which has been veryfied by extensive numerical 
experiments suggesting a physical origin of the phenomenon. 
In section \ref{linevolved}, we shall argue in
favour of the latter by presenting a linear stability analysis providing
an instability with appropriate frequencies and growth rates.

\subsection{Stability analysis of a model containing a captured shock}
\label{linevolved}

In this Section we shall initially assume and then prove a posteriori, that
the secondary oscillations of the 
captured shock described in section \ref{evolution} are caused by physical
processes. 
{
  We perform a linear stability analysis of a background model by assuming
  that the dependent variables radius, pressure, temperature and luminosity
  may be expressed as the sum of a background contribution and a small
  perturbation:
  \begin{eqnarray}
    x(m,t)=x_0(m,t) + x_1(m,t) \quad \rm{for}\quad x\in \{r,p,T,L\} 
  \end{eqnarray}
  The background coefficients $x_0(m,t)$ may be regarded as time independent, i.e.,
  $x_0(m,t)=x_0(m)$, 
  as long as the perturbations vary on much shorter timescales as the 
  background, i.e., as long as the condition
  \begin{eqnarray}
    \frac{d\log x_0(m,t)}{dt} \ll \frac{d\log x_1(m,t)}{dt} 
  \end{eqnarray}
  holds. $\frac{d}{dt}$ denotes the Lagrangian time derivative. 
Thus, the variations on dynamical timescales of the model containing 
the captured shock are regarded as stationary with respect to the
anticipated much faster instability.
Eigenmodes with periods of the order of the dynamical timescale
suffer from the competition with the variation of the ``background'' model,
whereas the approximation holds for those with much shorter periods.
For the model considered, the approximation is correct for $\abs{\sigma}>100$.
Should unstable eigenmodes of this kind exist, this would prove the 
instability and the high frequency oscillations of the captured shocks 
to be of physical origin.
Therefore, the results of a linear stability analysis of such 
a model are meaningful, as long as the obtained frequencies are interpreted
properly. 
}

A problem with this strategy is that the numerical simulations provide only
the superposition of the slow dynamical and the secondary, fast oscillations.
The linear stability analysis, however, requires a - on the fast oscillations - 
stationary background model. It is obtained by an appropriate time average over
a numerically determined sequence of models. ``Appropriate'' means, that the
average has to be taken over times longer than the short period 
oscillations and shorter
than the dynamical timescale. Thus all physical quantities 
$Q(m,t)$ are averaged according to 
\begin{eqnarray}
  <\!Q(m)\!>\ = \frac{1}{t_e-t_s}\int_{t_s}^{t_e} Q(m,t) {\rm{d}}t
\end{eqnarray}
where $t_s$ and $t_e$ are the beginning and the end of the averaging intervall and
satisfy the requirements discussed above.
$t_s$ has been varied between $4\cdot 10^7$ sec and $5\cdot 10^7$ sec 
(after the formation of the shock front) and the averaging interval between
$5\cdot 10^5$ sec and $1\cdot 10^6$sec.
All averages exhibit qualitatively
the same behaviour and the LNA stability analysis is largely
independent of the averaging parameters. The results presented in the following
were obtained for $t_s=5\cdot 10^7$ sec and $t_e=5.05\cdot 10^7$ sec.

\begin{figure*}
  \epsfig{file=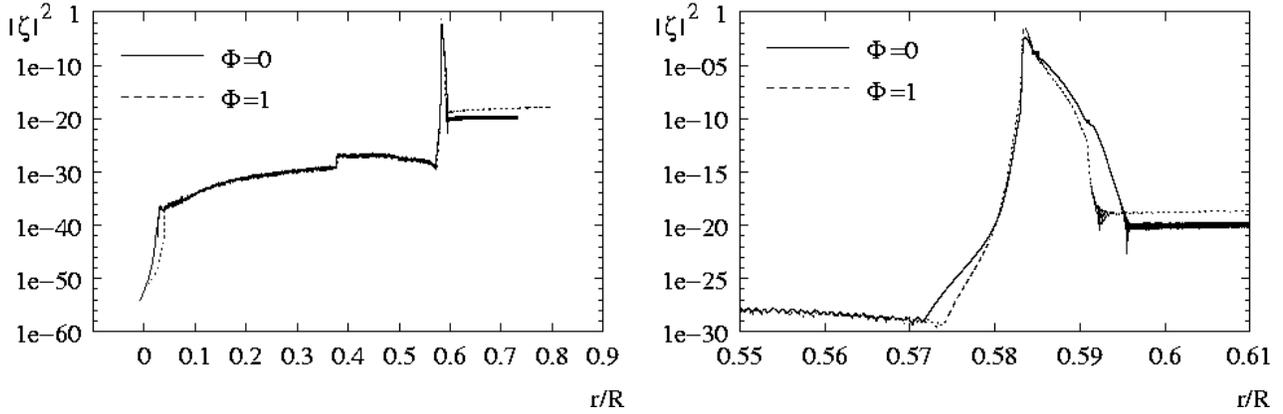,width=16.8cm}
  \caption{{Modulus of the Lagrangian displacement
    $\zeta$ as a function of relative radius for the eigenfrequencies  
    $\sigma_r=4651.7$, $\sigma_i=-119.6$ ($\Phi=0$, solid lines) and 
    $\sigma_r=4837.2$, $\sigma_i=-83.9$ ($\Phi=1$, dotted lines). 
    The right panel shows details
    for the shock zone.}} 
  \label{EigeHigh}
\end{figure*}

{
With these assumptions, the linear perturbation equations 
\ref{perturbation1}-\ref{perturbation4}, which have been derived
for a strictly static background model, remain valid even for the situation
studied here, except for
the momentum equation \ref{perturbation3}, which has to be modified according to:
\begin{eqnarray}
  p' = -(4+C_3\sigma^2)\zeta - Q_1 p \qquad {\rm{with}} \qquad
  Q_1=-\frac{\partial p_0}{\partial m}\frac{4\pi r_0^4}{Gm}
  \label{Q0}
\end{eqnarray}
$Q_1\neq 1$ accounts for the deviations from hydrostatic equilibrium.

As a result of the linear stability analysis (with $Q_1\neq 1$),
the expected unstable modes having high
frequencies have been identified. E.g., a typical mode of this kind satisfying
the assumptions discussed  
has the frequency
$\sigma_r=4837.2$ and the growth rate $\sigma_i=-83.9$.

In a second step, we investigate the influence of deviations from hydrostatic
equilibrium, i.e., the deviations from $Q_1=1$. For this purpose we rewrite 
equation \ref{Q0} as 
\begin{eqnarray} 
  p' = -(4+C_3\sigma^2)\zeta -  p + \Phi(1-Q_1)p 
\end{eqnarray}
with $0\le \Phi\le 1$.
The limits $\Phi=0$ and $\Phi=1$ correspond to hydrostatic
equilibrium and the averaged model containing the shock, respectively. 
The influence of deviations from hydrostatic
equilibrium is then studied by varying $\Phi$ between 0 and 1. 
Following the mode having 
$\sigma_r=4837.2$ and $\sigma_i=-83.9$ at $\Phi=1$ to $\Phi=0$ its frequency
and growth rate changes to $\sigma_r=4651.7$ and $\sigma_i=-119.6$. 
The moduli of the corresponding
Lagrangian displacements, which indicate the kinetic 
energy of the pulsations, are shown in Figure \ref{EigeHigh} as a function of
relative radius. The energy of the pulsation is concentrated 
around $r/R\approx 0.58$ and drops off exponentially above and below.
As a result,
neither eigenvalues nor eigenfunctions differ significantly for
$\Phi=0$ and $\Phi=1$, i.e., the assumption of hydrostatic equilibrium
is justified for the unstable modes considered. Therefore, we will assume
hydrostatic equilibrium in a further discussion and investigation of 
the secondary instability, i.e., all subsequent results
were obtained assuming $\Phi=0$.
}

\begin{table}
  \caption{Unstable modes of the averaged model}
  \label{mean}
  \begin{tabular}{crrrrrrrrrr}
    \hline
    $\sigma_r$ & 0.92  & 2.22  & 3.31 & 4.74  & 6.08   & 7.48  & 8.79 
    & 10.1  & 11.4   & 140.4 \\
    $\sigma_i$ & -0.05 & -0.33 & -0.32 & -0.25 & -0.29  & -0.18 & -0.21 
    & -0.24 & -0.12  & -0.1\\ 
    \hline    
    $\sigma_r$& 154.4& 162.9  & 157.2& 179.6 
    & 202.0 & 413.6 & 447.3  & 487.5 & 541.4 & 4651.7\\
    $\sigma_i$& -0.05 & -35.4 &-0.08& -27.8 
    & -17.6 & -30.4 & -26.3  & -54.9 &-0.35 & -119.6 \\
   \hline
  \end{tabular}

  \medskip
   $\sigma_r$  denotes the real part and  $\sigma_i$ the imaginary part of the 
   eigenfrequency $\sigma$ normalized by the global free fall time.
\end{table}

The results of a LNA stability analysis 
{{
according to equations 
\ref{perturbation1}-\ref{perturbation4} and Section \ref{LNA}
}}
for the averaged model are  summarized in Table \ref{mean}, where
representative values for the eigenfrequencies of unstable modes are given.
Three sets of unstable modes may be distinguished. 
Low order modes with $\sigma_r$
between $0.9$ and $9$ have growth rates of the order of $0.2$, i.e a ratio of 
$\frac{\sigma_i}{\sigma_r}\approx 0.1$. They can be identified with the
primary instability. However, their periods compete with the variation of 
the background model and therefore these modes have to be interpreted with caution.
The properties of two classes of high order unstable modes with frequencies 
$\sigma_r$ between $140$ 
{
and $4650$ are in accordance with our approximation.
}
One of them has high growth rates with  a ratio of 
$\frac{\sigma_i}{\sigma_r}\approx 0.1$,
the second low growth rates with a ratio of
$\frac{\sigma_i}{\sigma_r}\approx 5\cdot 10 ^{-4}$. The latter may be identified
with high order primary instabilities, whereas the former are attractive candidates
for the secondary, shock front instabilities sought.

\begin{figure*}
  \epsfig{file=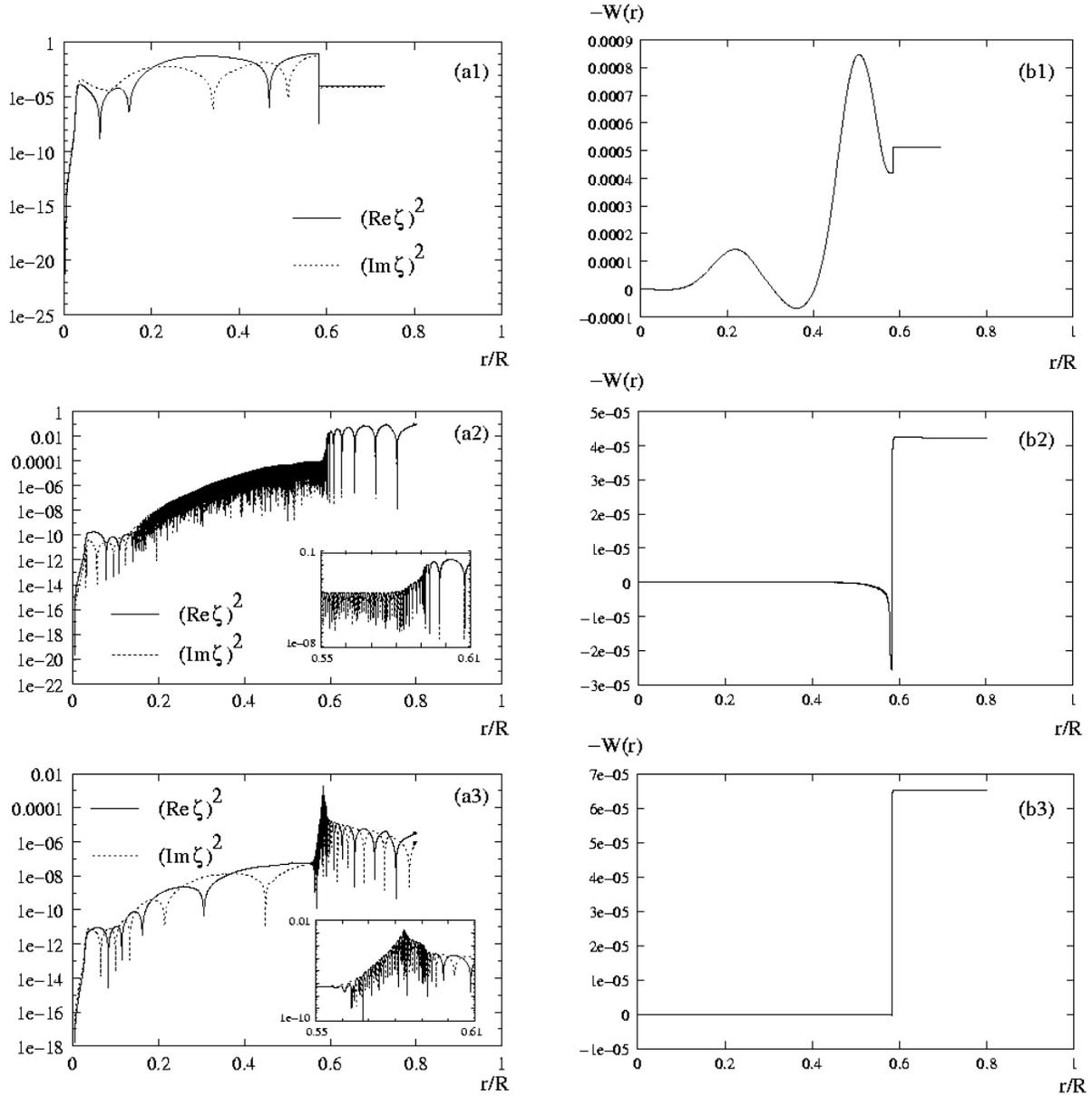,width=16.8cm}
  \caption{Lagrangian displacements $\zeta$ (a) and integrated workintegrals (b) 
    as a function of relative radius 
    for the  eigenfrequencies 
    $\sigma_r=2.22$, $\sigma_i=-0.33$ (1),
    $\sigma_r=157.2$, $\sigma_i=-0.08$ (2) and
    $\sigma_r=162.9$, $\sigma_i=-35.4$ (3) of the averaged model.}
  \label{eige.work}
\end{figure*}

For further discussion, we consider
eigenfunctions and the corresponding work integrals of 
representative members of the different sets of modes.
The work integral is a widely used tool to identify the regions in a star, which
drive or damp the pulsation. \citet{G94} has shown, that the concept of the 
work integral is not necessarily restricted to small values of 
the damping or growth rate. 
By replacing the conventional time average by an ensemble average it can 
be extended to arbitrary values of $\frac{\sigma_i}{\sigma_r}$. 
In any case, one arrives at the expression
\begin{eqnarray}
  W(r)=\frac{\sigma_i}{\sigma_r}\sim\int_0^r\pi {r'}^{2}\overline{p} \ {\rm{Im}}
  \left({ p^\dagger}
  {\rho^{\dagger\ast}}\right){\rm d}r'
  \label{work1}
\end{eqnarray}
for the work integral, where ${\rm{Im}}(z)$ denotes the imaginary
part of $z$, $()^*$ denotes complex conjugation and 
$p^\dagger$, $\rho^\dagger$ denote
the spatial parts of the eigenfunctions of the 
relative pressure and density perturbations, 
respectively. $\overline{p}$ is the pressure of the background model.
The sign of the integrand in equation \ref{work1} determines, if a region of 
the star damps or drives the pulsation, where
${\rm{Im}} ({ p^\dagger} {\rho^{\dagger\ast}})<0$ corresponds to driving
and ${\rm{Im}} ({ p^\dagger} {\rho^{\dagger\ast}})>0$ to damping.
Some authors (e.g. BKA) use $\log p$ instead of $r$ as independent variable
and therefore obtain an opposite sign of the differential work integral
for driving and damping influence.
To match this convention, $-W(r)$ is shown in figures
\ref{eige.work}.b, i.e., driving regions correspond to positive $-W(r)$, 
damping regions to negative $-W(r)$.

According to its eigenvalues the mode corresponding to $\sigma_r=2.22$, 
$\sigma_i=-0.33$ was identified as a primary instability. This is supported by the
Lagrangian displacement component $\zeta$ of the eigenfunction 
and the work integral shown in figure \ref{eige.work}.a1 and 
\ref{eige.work}.b1.
The shock front acts as an acoustic barrier causing the eigenfunction to 
vanish above it (figure \ref{eige.work}.a1).
The work integral (figure \ref{eige.work}.b1) exhibits two driving regions which
coincide with the opacity peaks at $\log T=5.3$ (caused by the contributions of 
heavy elements) and $\log T=4.7$ (He-ionization). The stability properties are not
affected significantly by the shock region. 

The counterparts of figures \ref{eige.work}.a1 and \ref{eige.work}.b1 for a 
weakly unstable high frequency mode having $\sigma_r=157.2$, $\sigma_i=-0.08$
are shown in figures \ref{eige.work}.a2 and \ref{eige.work}.b2. 
Again, the isolating effect of the shock front causes the dramatic variation of the
amplitude around $r/R_0=0.6$. However, in contrast to the 
eigenfunction presented in figure
\ref{eige.work}.a1 the amplitude is now significant above and negligible below the 
shock. High order modes of this kind in general exhibit strong damping. For the mode
considered the shock efficiently screens the inner damping 
part of the stellar envelope.
Thus the region below the shock contributes only weak damping which is overcome by 
the driving influence of the shock as shown by the work integral in figure
\ref{eige.work}.b2.

Apart from splitting the acoustic spectrum by an acoustic barrier 
into two sets of modes associated with the 
acoustic cavities below and above the shock, respectively, 
the shock itself gives rise 
to a third set. Lagrangian displacement and work integral for a typical member of
this set having $\sigma_r=162.9$, $\sigma_i=-35.4$ are shown in figures 
\ref{eige.work}.a3 and \ref{eige.work}.b3.
The amplitude of this unstable mode reaches its maximum on the shock and drops off 
exponentially above and below. 
Note its oscillatory behaviour on and close confinement to the shock.
The real parts of this set of eigenfrequencies of $\approx 200-500$ correspond
to periods of $\Pi\approx 8\cdot 10^4- 2\cdot 10^5$ sec, which are observed in the
luminosity perturbations (cf. figure \ref{Time}.c2) induced by the shock oscillations.
The work integral (figure \ref{eige.work}.b3) shows, that the shock is driving this
instability, and that the regions above and below do not contribute. 
Moreover, the basic assumption of stationarity of the averaged model holds for the 
frequencies and growth rates obtained.

Thus we have identified an instability by linear analysis of an averaged model, which
resembles the shock oscillations observed in the numerical simulations, both with 
respect to timescales and spatial structure. 
We therefore conclude, that the shock oscillations are not numerical artifacts. 
Rather they have a physical origin and are caused by an instability whose mechanism
will be investigated in detail in the following sections.
 
\subsection{Approximations}

In order to gain further insight into the physical processes responsible for the 
instability, different approximations in equations 
(\ref{perturbation1})-(\ref{perturbation4}) have been considered. 
To obtain a continuous transition from the exact treatment to the approximation, we
introduce a parameter $\Phi$ with $\Phi=1$ corresponding to the exact problem and 
$\Phi \rightarrow 0,\infty$ to the approximation. The numerical results, i.e. the 
eigenvalues of the shock front instabilities, are followed as a function of $\Phi$.

Introducing $\Phi$ into the Euler equation as
\begin{eqnarray}
  p' = -(4+\Phi\cdot C_3\sigma^2)\zeta -p 
\end{eqnarray}
the limit $\Phi\rightarrow 0$ corresponds to vanishing acceleration
and implies the elimination of acoustic modes from the spectrum, which then
only consists of secular modes. Application of this limit to the shock instabilities
has not revealed any unstable modes. Rather the eigenvalues have diverged.
This excludes a thermal origin of both the unstable modes and the 
instability mechanism. For a proper treatment of the instability, the 
mechanical acceleration has to be taken into account.

Introducing $\Phi$ into the equation for energy conservation as
\begin{eqnarray}
  l'     = C_1\cdot\Phi\cdot(i\sigma)(-p+ C_2 t)
\end{eqnarray}
the adiabatic limit is obtained by $\Phi\rightarrow \infty$.
The latter implies $(-p+ C_2 t)=0$, i.e. the algebraic adiabatic relation 
between pressure and temperature perturbation. 
No unstable modes have been found following the shock instabilities
into the adiabatic limit.

Introducing $\Phi$ into the equation for energy conservation as 
\begin{eqnarray}
  l'     = C_1\cdot\Phi\cdot(i\sigma)(-p+ C_2 t)
\end{eqnarray}
$\Phi\rightarrow 0$ corresponds to the so called NAR-limit
(Non-Adiabatic-Reversible limit) \citep{GG90}. 
Although this approximation - like the adiabatic approximation - implies 
constant entropy, it does not represent the adiabatic limit  
$(-p+ C_2 t)=0$. Rather it is equivalent to $C_1 \rightarrow 0$. Since $C_1$
is related to the thermal and dynamical timescales $\tau_{th}$ and 
$\tau_{dyn}$ by
\begin{eqnarray}
C_1=\frac{\tau_{th}}{\tau_{dyn}} \nabla_{ad}\sqrt{\frac{C_3C_4}{\Gamma}}
\end{eqnarray}
this approximation is also being refered to as the zero thermal timescale
approximation ($\Gamma$ and $\nabla_{ad}$ are the adiabatic indices). 
Physically, it means that the specific heat of the envelope
is negligible and luminosity perturbations cannot be sustained. In particular,
this approximation rules out the classical $\kappa$-mechanism as the source of
an instability - should it exist in the NAR-limit -
since this Carnot-type process relies on a finite 
heat capacity. When following the frequencies of the modes belonging 
to the shock front instabilities into the NAR-limit, periods and growth rates 
change only slightly (by at most 10 per cent). Thus the NAR-approximation 
may be regarded as a satisfactory approximation and will form the basis of our
investigations in the following sections.

\section{AN ANALYTICAL MODEL}
\label{analyticalmodel}

\subsection{Three-Zone-Model}
\label{threezonemodel}

The modal structure identified in section \ref{linevolved} with three sets of modes
associated with three acoustic cavities (inner envelope, shock and outer envelope)
suggests the construction of a three zone model. In order to enable an analytical 
solution, the coefficients of the differential equations are kept constant in each
zone.  

According to the previous section the NAR-approximatin is sufficient to describe
the shock front instabilities. The equation of energy conservation is then satisfied
identically and luminosity perturbations vanish. Thus we are left with a system of
third order comprising the mechanical equations and the diffusion equation with 
zero luminosity perturbation.

\begin{figure}
  \epsfig{file=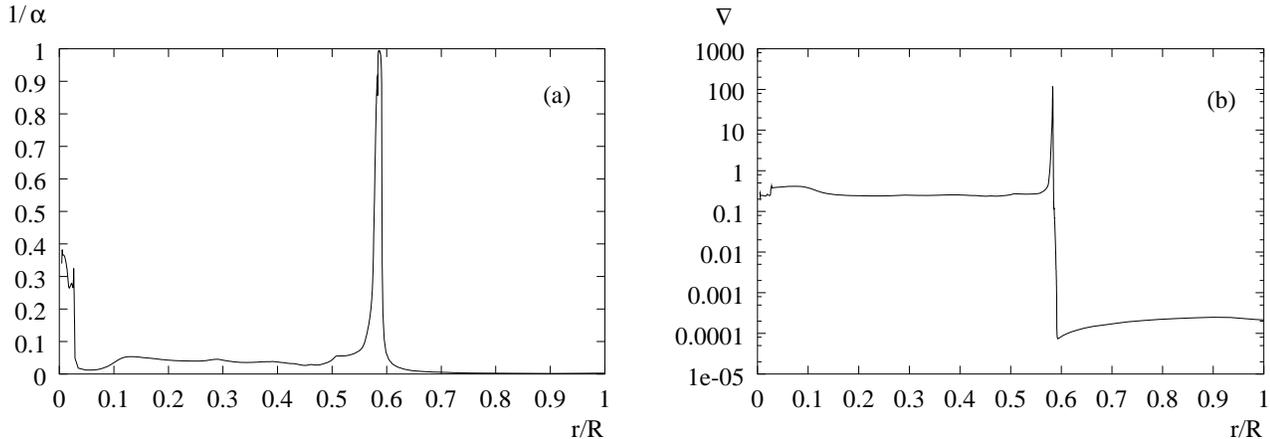,width=16.8cm}
  \caption{The coefficients $C_5=\alpha$ (a) and $C_7=\nabla$ (b)
    of the averaged model as a function of relative radius.} 
  \label{C7}   
\end{figure}

Further reduction of the order of the differential system is achieved by considering
its coefficients which depend on the properties of the averaged model. 
In figure \ref{C7} the coefficients 
$\left.C_5=\alpha=\frac{\partial\log \rho}{\partial\log p}\right|_T
\approx \frac{1}{\beta}$ and $C_7=\nabla=\frac{{\rm{d}}\log T}{{\rm{d}}\log p}$ 
are shown as a function of relative radius. $\beta$ denotes the ratio of gas pressure
to total pressure. The coefficients $C_4=-\frac{d \log r}{d \log p}$ and 
$C_3=\frac{4\pi r^3\overline{\rho}}{M_r}$ may be regarded as 
constant all over the envelope. Approximate values are
$C_4\approx\frac{1}{3}$ and $C_3\approx 3$. The latter holds because almost the 
entire mass is concentrated in the stellar core.
From figure \ref{C7} we deduce that radiation pressure is dominant except for the
shock zone. Therefore we replace the diffusion equation \ref{perturbation4} by the 
algebraic equation of state for pure radiation ($p=4t$) in the inner 
and outer envelope. On the other hand $C_7=\nabla$ can - to first approximation -
be regarded as singular in the 
shock zone. According to equation \ref{perturbation4} this requires the expression
$(-4\zeta+C_8p-C_9t)$ to vanish there. Thus the differential diffusion equation
is replaced by an algebraic relation in all three zones, reducing the system to 
second order.

Adopting the alternative notation \citep{BK62} $C_6=\delta$, 
$C_8=\kappa_p$ and $C_9=4-\kappa_T$, where $\delta$ is the negative 
logarithmic derivative of density with respect to temperature at constant pressure,
$\kappa_p$ the logarithmic derivative of opacity with respect to pressure at 
constant temperature and $\kappa_T$ the logarithmic derivative of 
opacity with respect 
to temperature at constant pressure, and choosing the relative radius $x$ as the 
independent variable, we are left with the following set of equations: 
\begin{eqnarray}
  \frac{1}{\psi}\frac{d\zeta}{dx}     &=&\frac{1}{3}(3\zeta +\alpha p-\delta t) \\
  \frac{1}{\psi}\frac{dp}{dx}         &=& -(4+3\sigma^2)\zeta -p \\
  t          &=&
  \begin{cases}
    \left(\frac{1}{4-\kappa_T}\right)(\kappa_p p -4\zeta)
    \quad & x\in[a,b]\\
    \frac{1}{4} \cdot p  \ &x\in[0,a) \ \mbox{or} \ x\in(b,1]
  \end{cases}
\end{eqnarray}
$a$ and $b$ denote the lower and upper boundary of the shock zone. The transformation
of the independent variables $\ln p_0 \rightarrow x$ introduces the factor $\psi$,
which is constant within the framework of the three-zone-model, and 
given by an appropriate
mean of the quantity $\frac{-1}{C_4 x}$. In general $\psi$ is negative and
of order unity.

We are thus left with a system of second order consisting
of the mechanical equations (continuity and Euler equations) which is closed 
by the algebraic relations $0=-4\zeta+C_8p-C_9t$ and $p=4\cdot t$ for the shock region
and the inner/outer regions, respectively.
We rewrite it as:
\begin{eqnarray}
  \frac{d\zeta}{dx}&=&A_{1,2}\cdot\zeta+B_{1,2}\cdot p \label{system1}\\
  \frac{dp}{dx}&=& C\cdot \zeta +D\cdot p
  \label{system2}
\end{eqnarray}
where
\begin{eqnarray}
  &A_{1,2}&=
  \begin{cases}
    \psi\left(1+\frac{\frac{4\delta}{3}}{4-\kappa_T}\right) \ &x\in[a,b]\\
    \psi &x\in[0,a) \ \mbox{or} \ x\in(b,1]
  \end{cases}\\
\\
  &B_{1,2}&= 
  \begin{cases}\psi\left(\frac{\alpha}{3}-
      \frac{\frac{\delta\kappa_p}{3}}{4-\kappa_T}\right) 
    \ &x\in[a,b]\\
     \psi\left(\frac{\alpha}{3}-\frac{\delta}{12}\right)&x\in[0,a) 
     \ \mbox{or} \ x\in(b,1]
  \end{cases} \\
\\
  &C&=-\psi(4+3\sigma^2) \\
  &D&=-\psi
\end{eqnarray}
and the subscript $1$ denotes the values of the coefficients in the shock region,
the subscript $2$ values in the inner and outer regions.
We introduce new variables by
\begin{eqnarray}
&\hat{\zeta}&=e^{\int A_{1,2} dx}\cdot \zeta\\
&\hat{p}&=e^{\int D dx} \cdot p
\end{eqnarray}
The system \ref{system1}-\ref{system2} then reads
\begin{eqnarray}
  \frac{d\hat{\zeta}}{dx}&=&
  B_{1,2}\cdot\hat{p}\cdot e^{\int D dx}\cdot e^{-\int A_{1,2} dx} \label{e1}\\
  \frac{d\hat{p}}{dx}&=&
  C\cdot\hat{\zeta}\cdot e^{-\int D dx}\cdot e^{\int A_{1,2} dx}\label{e2}
\end{eqnarray}
These equations are equivalent to the following single second order equation:
\begin{eqnarray}
  \frac{d}{dx} 
  \left(
    \frac{1}{C}\cdot e^{\int D dx}\cdot 
    e^{-\int A_{1,2} dx}\cdot\frac{d\hat{p}}{dx}
  \right)
  -  e^{\int D dx}e^{-\int A_{1,2} dx}\cdot B_{1,2}\cdot \hat{p}=0
  \label{secondorder}
\end{eqnarray}

\subsubsection{Mathematical Structure of the Problem}
\label{math}

Equation \ref{secondorder} may be written as
\begin{eqnarray}
  \frac{d}{dx} 
  \left(
     e^{\int D dx}\cdot 
    e^{-\int A_{1,2} dx}\cdot\frac{d\hat{p}}{dx}
  \right)
  +4\cdot\psi\cdot  e^{\int D dx}e^{-\int A_{1,2} dx}\cdot B_{1,2}\cdot \hat{p}
  +3\sigma^2\cdot\psi\cdot  e^{\int D dx}e^{-\int A_{1,2} dx}\cdot B_{1,2}\cdot
  \hat{p} = 0
  \label{sturm0}
\end{eqnarray}
and has the form 
\begin{eqnarray}
  \frac{d}{dx}\left[q(x)\frac{d}{dx}\hat{p}\right] 
  -w(x)\hat{p} + \lambda u(x)\hat{p}=0
\label{sturm}
\end{eqnarray}                          
with $q(x)>0$ in the integration intervall. However, $u(x)$ is positive
in the inner and outer regions and negative in the shock region, i.e.,
$u(x)$ changes sign in the integration intervall. (This holds also for $w(x)$.)
Therefore, this problem is not of 
Sturm-Liouville type. 
On the other hand, if we consider each zone separately with boundary conditions
$\hat{p}=0$, equation \ref{sturm} describes a Sturm-Liouville
problem. 
In the shock zone we define eigenvalues $\lambda=-\sigma^2$ and thus have 
$u(x)>0$, $w(x)>0$, for the inner and outer
zones we get $u(x)>0$, $w(x)<0$ by defining $\lambda=\sigma^2$.

For a Sturm-Liouville problem, the eigenvalues are real and form a sequence 
\begin{eqnarray}
\lambda_1 < \lambda_2 < \lambda_3 < \lambda_4 < \ldots
\end{eqnarray} 
Furthermore, $\lambda_1$ may be estimated on the basis of  
the variational principle
\begin{eqnarray}
  \lambda_1=\min_{\hat{p}}
  \frac{\int_0^1 q(x)\abs{\hat{p}'}^2dx+\int_0^1w(x)\abs{\hat{p}}^2dx}
  {\int_0^1 u(x)\abs{\hat{p}}^2dx}
\end{eqnarray} 

Therefore $\lambda_1=-\sigma^2_1$ is positive in the shock zone, since $w(x)$ 
is positive there. This means that 
we have purely imaginary eigenfrequencies $\sigma_j=\pm i\sqrt{\lambda_j}$ with 
positive $\lambda_j$ and 
\begin{eqnarray}
\abs{\sigma_{1}} < \abs{\sigma_{2}} < \abs{\sigma_{3}} 
< \abs{\sigma_{4}} < \ldots
\end{eqnarray} 
Thus the shock zone provides unstable eigenfrequencies.

Since $w(x)$ is negative in the inner and outer zones, we cannot guarantee
$\lambda_1$ to be positive there. For sufficiently large $j$, however, $\lambda_j$ will
always become positive. As a consequence, all eigenfrequencies 
$\sigma_j=\pm \sqrt{\lambda_j}$ will become real
for sufficiently high order $j\ge n$ and satisfy:
\begin{eqnarray}
  \abs{\sigma_{n}} < \abs{\sigma_{{n+1}}} < \abs{\sigma_{{n+2}}} 
< \abs{\sigma_{{n+3}}} < \ldots
\end{eqnarray} 

In principle, the mathematical structure of the problem allows for 
imaginary pairs of eigenfrequencies at low orders in the inner and outer zones. 
For the particular parameters studied in the following sections, however, 
$\lambda_1$ turned out to be positive, i.e., $n=1$ and all eigenfrequencies are
real. 

Even if equation \ref{sturm0} together with the boundary conditions
$\hat{p}=0$ at $x=0$ and $x=1$ (three-zone-model) is not of Sturm-Liouville type,
the differential operator
\begin{eqnarray}
\mathcal{D}=\frac{d}{dx}\left(q\frac{d}{dx}\right)+\lambda u-w
\end{eqnarray} 
in equation  \ref{sturm} can be shown to be self adjoint with the boundary conditions
$\hat{p}=0$ at $x=0$ and $x=1$.
Therefore the eigenvalues $\lambda$ are real and we do expect only real or  purely
imaginary eigenfrequencies $\sigma$, i.e., we will not be able to reproduce
the complex eigenfrequencies of the exact problem in this approximation. 

\subsubsection{Results}

Assuming the coefficients $C$ and $B_{1,2}$ to be constant,
equations \ref{e1} and \ref{e2} are solved by the Ansatz 
$\hat{p},\hat{\zeta}\propto e^{k x}$. For the wavenumbers $k$
we get 
\begin{eqnarray}
k=\pm\sqrt{B_{1,2}\cdot C}
\end{eqnarray}
Thus the general solutions reads
\begin{eqnarray} 
  &\hat{p}=
  \begin{cases}
    a_1\cdot e^{\sqrt{B_2C}x} + a_2\cdot e^{-\sqrt{B_2C}x}  
    &\quad x\in[0,a) \\
    b_1\cdot e^{\sqrt{B_1C}x} + b_2\cdot e^{-\sqrt{B_1C}x}
    &\quad x\in[a,b] \\
    c_1\cdot e^{\sqrt{B_2C}x} + c_2\cdot e^{-\sqrt{B_2C}x}
    &\quad x\in(b,1] \\
  \end{cases}
\end{eqnarray}
$a_{1,2}$, $b_{1,2}$ and $c_{1,2}$ are integration constants and have
to be determined by
the requirements of continuity and differentiability of $\hat{p}$ at $x=a$
and $x=b$ and the boundary conditions at $x=0$ and $x=1$. For the latter we choose
$\hat{p}=0$, which implies
\begin{eqnarray}
&a_2&=-a_1 \qquad \qquad\qquad \mbox{and}\\
&c_2&=-c_1\cdot e^{2\sqrt{B_2C}}
\end{eqnarray}
Together with the requirements of continuity and differentiability this yields 
the dispersion relation 
\begin{eqnarray}
  \frac{(\sqrt{B_1C}-\sqrt{B_2C})e^{\sqrt{B_2C}a}+
    (-\sqrt{B_1C}-\sqrt{B_2C})e^{-\sqrt{B_2C}a}}
  {(\sqrt{B_1C}+\sqrt{B_2C})e^{\sqrt{B_2C}a}+
    (-\sqrt{B_1C}+\sqrt{B_2C})e^{-\sqrt{B_2C}a}}
  \cdot 
  e^{2\sqrt{B_1C}(a-b)} = \notag \\
  \frac  {(\sqrt{B_1C}-\sqrt{B_2C})e^{\sqrt{B_2C}(b-1)}+
    (-\sqrt{B_1C}-\sqrt{B_2C})e^{-\sqrt{B_2C}(b-1)}}
  {(\sqrt{B_1C}+\sqrt{B_2C})e^{\sqrt{B_2C}(b-1)}+
    (-\sqrt{B_1C}+\sqrt{B_2C})e^{-\sqrt{B_2C}(b-1)}}
  \label{dispersion}
\end{eqnarray}
where the eigenfrequencies $\sigma$ are contained in the coefficient $C$. In general,
the roots of equation \ref{dispersion}
have to be calculated numerically, using, for example,
a complex secant method. Separate spectra for the three 
isolated zones may be obtained by 
assuming the boundary conditions $\hat{p}=0$ at $x=a,b$ instead of 
continuity and differentiability requirements. We are then left with the 
dispersion relations
\begin{eqnarray}
&D_{inner}&=1=   \frac{(\sqrt{B_1C}-\sqrt{B_2C})e^{\sqrt{B_2C}a}+
    (-\sqrt{B_1C}-\sqrt{B_2C})e^{-\sqrt{B_2C}a}}
  {(\sqrt{B_1C}+\sqrt{B_2C})e^{\sqrt{B_2C}a}+
    (-\sqrt{B_1C}+\sqrt{B_2C})e^{-\sqrt{B_2C}a}}\label{D1}\\
  &D_{outer}&=1=  \frac
  {(\sqrt{B_1C}+\sqrt{B_2C})e^{\sqrt{B_2C}(b-1)}+
    (-\sqrt{B_1C}+\sqrt{B_2C})e^{-\sqrt{B_2C}(b-1)}}
  {(\sqrt{B_1C}-\sqrt{B_2C})e^{\sqrt{B_2C}(b-1)}+
    (-\sqrt{B_1C}-\sqrt{B_2C})e^{-\sqrt{B_2C}(b-1)}} \label{D2}\\
  &D_{shock}&=1= e^{2\sqrt{B_1C}(a-b)} \label{D3}
\end{eqnarray}
for the inner, outer and shock zones, respectively.

For the averaged model we have $B_1\approx -4\psi$ and dominant radiation 
pressure implies $B_2\approx \frac{1}{4}\psi$. Inserting these values into 
equations \ref{D1}-\ref{D2} we are left with
\begin{eqnarray}
 & \sqrt{\frac{\psi^2}{4}(4+3\sigma^2)}a&=\frac{(2n+1)\pi}{2}\qquad n \in \mathbb{Z}
  \label{sigma1}\\
 & \sqrt{\frac{\psi^2}{4}(4+3\sigma^2)}(b-1)&=\frac{(2n+1)\pi}{2}\qquad n 
 \in \mathbb{Z}
  \label{sigma2}
\end{eqnarray}
Thus we have real $\sigma$, i.e., neutrally stable modes, if the inner and outer 
regions are considered separately, in accordance with the discussion in 
section \ref{math}. For the shock region equation \ref{D3} yields 
\begin{eqnarray} 
  2\sqrt{4\psi^2(4+3\sigma^2})(b-a)
  =2\pi n i\qquad n \in \mathbb{Z}
\label{sigma3}
\end{eqnarray}
These solutions correspond to purely imaginary $\sigma$ implying instability.
The solutions of equations \ref{sigma1}-\ref{sigma3} 
can be used as initial guesses for the 
numerical iteration of equation \ref{dispersion}, the disperion relation of the 
three-zone-model. Some representative eigenvalues of the three-zone-model 
are given in Table \ref{roots2}.
\begin{table}              
  \caption{Eigenfrequencies $\sigma$ ($\sigma_r$: real part, $\sigma_i$: imaginary 
    part) of the three-zone-model having the parameters
    $B_1=-4\psi$, $B_2=\frac{\psi}{4}$, $a=0.57$, $b=0.59$, $\psi=-1$}
  \label{roots2}
  \begin{tabular}{crrrrrrr}
    \hline
    $\sigma_r$ & 12.01  & 16.78  & 18.55  & 24.86  & 25.83   & 31.29  & 34.67 \\
    $\sigma_i$ & 0      & 0      & 0      &  0     &  0      &  0     & 0\\  
    \hline
    $\sigma_r$ & 0    &   0   &  0   &   0    &   0\\
    $\sigma_i$ & 7.21 & 52.43 &97.77 & 143.11 & 188.46\\
    \hline
  \end{tabular}
\end{table}
 
Once the eigenfrequencies are determined, the corresponding 
eigenfunctions are given by
\begin{eqnarray} 
  &{p}=
  \begin{cases} a_1\cdot e^{\psi x}\cdot e^{\sqrt{B_2C}x} + 
    a_2\cdot e^{\psi x}\cdot e^{-\sqrt{B_2C}x}  
    &\quad x\in[0,a) \\
    b_1\cdot e^{\psi x}\cdot e^{\sqrt{B_1C}x} + b_2\cdot e^{\psi x}\cdot e^{-\sqrt{B_1C}x}
    &\quad x\in[a,b] \\
    c_1\cdot e^{\psi x}\cdot e^{\sqrt{B_2C}x} + c_2\cdot e^{\psi x}\cdot  e^{-\sqrt{B_2C}x}
    &\quad x\in(b,1] \\
  \end{cases}
\end{eqnarray}
The factor $e^{\psi x}$ is due to the transformation from $p$ to $\hat{p}$.

\begin{figure}
  \epsfig{file=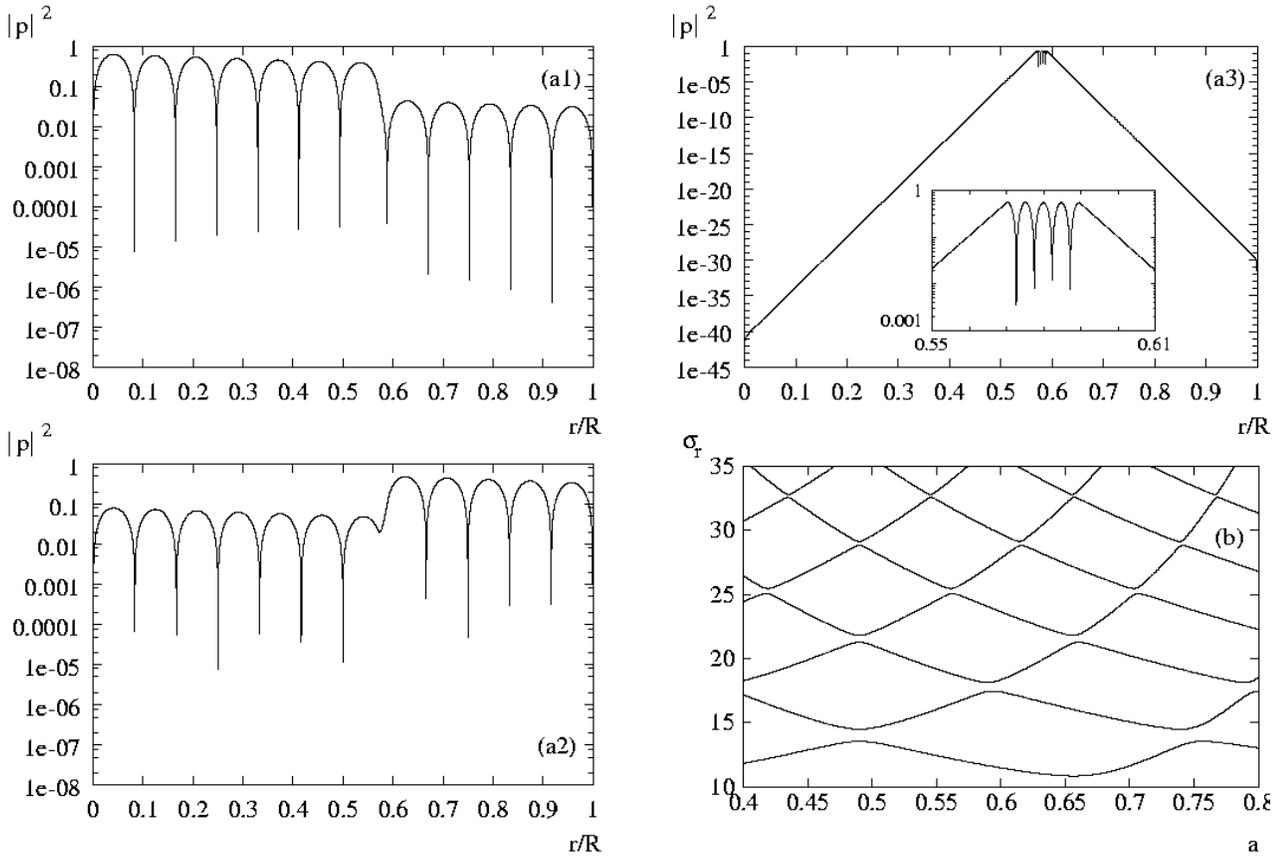,width=16.8cm}
  \caption{Eigenfunctions for the three-zone-model with the parameters
    $B_1=-4\psi$, $B_2=\frac{\psi}{4}$, $a=0.57$, $b=0.59$, $\psi=-1$,
    and the frequencies $\sigma_r=44.04$, 
    $\sigma_i=0$ (a1), 
    $\sigma_r=43.53$, $\sigma_i=0$ (a2), 
    $\sigma_r=0$, $\sigma_i=188.46$ (a3). (b): Eigenfrequencies $\sigma_r$ of 
    neutrally stable modes as a function of the position $a$ of the lower boundary
    of the shock region for fixed $b-a=0.02$ and  
    $B_1=-4\psi$, $B_2=\frac{\psi}{4}$, $\psi=-1$.}
  \label{analytisch}   
\end{figure}

Typical eigenfunctions are presented in figures 
\ref{analytisch}.a1-\ref{analytisch}.a3.
Three types of modes may be distinguished belonging to the three zones of
the model. Real eigenfrequencies are associated with the inner and outer region.
Except for the shock region they are oscillatory and reach their maximum in
the respective region. ``Shock modes'' correspond to unstable and damped modes
(purely imaginary pairs of eigenvalues). They oscillate in the shock region and are
evanescent elsewhere. We note the correspondence of figures \ref{analytisch}.a1
and \ref{eige.work}.a1, \ref{analytisch}.a2 and \ref{eige.work}.a2 and 
\ref{analytisch}.a3 and \ref{eige.work}.a3, i.e., the results of the analytical
model resemble those of the exact analysis. 
The influence of the shock position on the modal structure may also be studied within 
the framework of the three-zone-model. As long as the width of the shock zone and
the coefficient $B_1$ are not varied, the ``shock modes'' are not affected. 
The dependence on the shock position of the neutrally stable ``inner'' and ``outer'' 
modes is shown in figure \ref{analytisch}.b. Moving the shock position outwards,
the frequencies of the inner modes decrease, whereas those of the outer modes 
increase, according to the variation of the length of the corresponding 
acoustic cavities. This leads inevitably to multiple crossings between the frequencies
of the inner and outer modes, which unfold into avoided crossings 
(see, e.g., \cite{GG90}). Mode interaction by instability bands is excluded here
according to the general discussion in section \ref{math}.

\subsubsection{Interpretation}
\label{intepret}

The three-zone-model reproduces the effects of the shock front
regarding important aspects: The front acts as an acoustically isolating layer
which separates the inner and outer part of the envelope. As a result, these parts
provide largely independent spectra. This may be illustrated by the variation of 
the position of the shock front. Apart from the expected spectra associated with 
the inner and outer envelope,
an additional spectrum of modes is generated by the shock region itself.

Comparing eigenfunctions of the averaged and the analytical model 
(figures \ref{eige.work}.a1-a3 and figures \ref{analytisch}.a1-a3), we find a strikingly
similar behaviour. 
In particular, the confinement of the unstable shock modes is present in
both cases.
Due to constant coefficients, however, the analytical model reproduces neither
decreasing amplitudes nor increasing spatial frequencies towards the stellar center. 

We have identified unstable modes in the shock zone of the analytical model.
They resemble those of the shock instabilities of the averaged model, 
and are related to the sound travel time across the shock zone.
Its radial extent is primarily responsible for their high frequencies.
 
The analysis in section \ref{math} has shown, that the sign of $u(x)$ in 
equation \ref{sturm} is responsible for the instability in the
shock region. This sign is determined by the term $\frac{B_1}{\psi}$, which is 
given by
\begin{eqnarray}
\frac{B_1}{\psi}=\frac{\alpha}{3}-\frac{\frac{\delta\kappa_p}{3}}{4-\kappa_T} 
\label{b1psi}
\end{eqnarray}
Estimating the various terms in equation \ref{b1psi}, we find that the sign of $\kappa_p$
determines the sign of $\frac{B_1}{\psi}$. 
A dependence on the sign of $\kappa_p$ of the instability, however, 
is not recovered in the exact problem, which can be 
tested by replacing $\kappa_p$ with $-\kappa_p$ there. 
The exact problem is not affected by this substitution. Thus we conclude, that
the analytical model does not provide correct results in this respect and needs to 
be refined to describe the instability properly. 
In order to investigate
the origin of the instability, some of the simplifying assumptions of the
analytical model need to be dropped. 
In this direction, a more realistic model of 
the shock zone will be presented in the following section.

\subsection{Shock-Zone-Model}

Our study of the three-zone-model in section \ref{threezonemodel} has shown, 
that inner, outer and shock zones may to good approximation be 
treated separately by assuming suitable
boundary conditions, e.g., vanishing pressure at boundaries and interfaces. 
Moreover, the instabilities of interest are not provided by the inner and outer zones.
Therefore we restrict the following study to the shock zone by applying 
the boundary conditions ${p}(a)={p}(b)=0$. Within the framework of the analytical 
model the coefficients of the perturbation equations are taken to be
constant with the values given in section  \ref{threezonemodel}.

Contrary to section \ref{threezonemodel} we will not replace the diffusion 
equation by an algebraic relation here, as this  turned out
to lead to erroneous results. However, we still adopt the NAR-approximation.
The set of equations considered then reads:
\begin{eqnarray}
\frac{1}{\psi}\frac{d\zeta}{dx}&=&\frac{1}{3}(3\zeta +\alpha p-\delta t)  \label{f1}\\
\frac{1}{\psi} \frac{d p}{dx}  &=& -(4+3\sigma^2)\zeta -p \label{f2}\\
\frac{1}{\psi} \frac{d t}{dx}   &=& \nabla(-4\zeta+\kappa_pp-(4-\kappa_T)t)\label{f3}\\ 
 \frac{1}{\psi}\frac{d l}{dx}         &=& 0\label{f4}
\end{eqnarray}
Written in matrix form this yields
\begin{eqnarray}
  \frac{1}{\psi}\frac{d}{dx}
  \begin{pmatrix}
    \zeta \\
    p \\ 
    t \\ 
    l
  \end{pmatrix}=
  \begin{pmatrix}
    1             & \frac{\alpha}{3} & -\frac{\delta}{3} & 0 \\
    -(4+3\sigma^2) &      -1         &     0             & 0 \\
    -4\nabla       & \nabla\kappa_p  & -\nabla(4-\kappa_T)& 0\\
      0            &  0              &     0             & 0\\
  \end{pmatrix}
  \begin{pmatrix}
    \zeta \\ p \\ t \\ l
  \end{pmatrix}
\end{eqnarray}
The differential equation is solved by an exponential dependence $\propto e^{ikx}$ of the 
dependent variables. 
Thus we arrive at the linear algebraic equation 
\begin{eqnarray}
  \begin{pmatrix}
    1 - \frac{ik}{\psi}            & \frac{\alpha}{3} & -\frac{\delta}{3} & 0 \\
    -(4+3\sigma^2) &      -1 -\frac{ik}{\psi}        &     0             & 0 \\
    -4\nabla       & \nabla\kappa_p  & -\nabla(4-\kappa_T)-\frac{ik}{\psi}& 0\\
     0            &  0              &     0             & -\frac{ik}{\psi}\\
  \end{pmatrix}
  \begin{pmatrix}
    \zeta \\ p \\ t \\ l
  \end{pmatrix}
  =
  \begin{pmatrix}
    0 \\ 0 \\ 0 \\ 0
  \end{pmatrix}
\end{eqnarray}
This equation has a non trivial solution only if the determinant of the matrix
vanishes, which provides a quartic equation for the wavenumber $k$. One of its roots
is zero, the remaining three roots are determined by the following cubic equation:
\begin{eqnarray}
  \left(\frac{ik}{\psi}\right)^3
  \frac{1}{\nabla (4-\kappa_T)} + \left(\frac{ik}{\psi}\right)^2 +
 \left(\frac{ik}{\psi}\right)
  \underbrace{\left(
    -\frac{\frac{4}{3}\delta}{(4-\kappa_T)} -\frac{1}{\nabla(4-\kappa_T)}
    (1-\frac{\alpha}{3}(4+3\sigma^2))
  \right)}_{d_1} +\notag\\
  \underbrace{\left(
    -1+\frac{\alpha}{3}(4+3\sigma^2)-\frac{1}{(4-\kappa_T)}
      \left(
        \frac{\delta}{3}\kappa_p(4+3\sigma^2) +\frac{4}{3}\delta
      \right)
  \right)}_{d_2}=0\label{cubic}
\end{eqnarray}
In the limit of large $\nabla$ they may be given in closed form:
\begin{eqnarray}
\left(\frac{ik}{\psi}\right)_{1,2}=-\frac{d_1}{2} \pm\sqrt{\frac{d_1^2}{4}-d_2}
\label{k1k2}
\end{eqnarray}
\begin{eqnarray}
\left(\frac{ik}{\psi}\right)_3=-\nabla(4-\kappa_T)
\end{eqnarray}
The general solution to the perturbation problem consists of a superposition
of four fundamental solutions associated with the four roots for the wavenumber, two
of which are oscillatory (those associated with $k_1$ and $k_2$).
The dispersion relation is then derived by imposing four conditions. In addition
to the boundary conditions $p=0$ at $x=a,b$, we require the two non oscillatory fundamental
solutions not to contribute to the eigensolution. The latter is then only determined by
$k_1$ and $k_2$:
\begin{eqnarray}
  p=h_1\cdot e^{ik_1x}+h_2\cdot e^{ik_2x}
\end{eqnarray}
where $h_1$ and $h_2$ are integration constants. They are determined by the
boundary conditions $p=0$ at $x=a,b$, which imply
\begin{eqnarray}
k_1-k_2=\frac{2\pi n}{(a-b)}
\end{eqnarray}
where $n \in \mathbb{Z}$ denotes the order of the overtone.
Using equation \ref{k1k2} we get
\begin{eqnarray}
\frac{d_1^2}{4}-d_2=-\frac{\pi^2 n^2}{\psi^2(a-b)^2}
\label{k1-k2}
\end{eqnarray}
With the definitions of $d_1$ and $d_2$ (equation \ref{cubic}) we arrive at a 
quadratic equation in $\sigma^2$.
Expanding the coefficients of $\sigma^2$ in terms of $\frac{1}{\nabla}$ and 
assuming $\frac{\pi^2n^2}{(a-b)^2}$ to be large, we obtain to lowest order in 
$\frac{1}{\nabla}$:
\begin{eqnarray}
\sigma^4\frac{\frac{9}{6}\alpha^2}{\nabla^2(4-\kappa_T)^2}
-\sigma^2\left(\alpha-\frac{\delta\kappa_p}{(4-\kappa_T)}
\right)
+\frac{\pi^2n^2}{\psi^2(a-b)^2}=0
\end{eqnarray} 
Defining 
\begin{eqnarray}
\tilde{\nabla}=\frac{\nabla^2(4-\kappa_T)^2}{\frac{9}{6}\alpha^2}
\end{eqnarray}
this equation has the solutions
\begin{eqnarray}
  \sigma_{1,2}^2=\frac{\tilde{\nabla}}{2}
  \left(\alpha-\frac{\delta\kappa_p}{(4-\kappa_T)}
  \right) \pm
  \frac{\tilde{\nabla}}{2}
  \left(\alpha-\frac{\delta\kappa_p}{(4-\kappa_T)}
  \right) 
  \sqrt{1-\frac{4}{\tilde{\nabla}}\frac{\pi^2n^2}{\psi^2(a-b)^2}
    \frac{1}{\left(\alpha-\frac{\delta\kappa_p}{(4-\kappa_T)}
  \right)^2}}
\label{letzte}
\end{eqnarray}
In the NAR-approximation, eigenfrequencies come in complex conjugate pairs, i.e.,
complex eigenfrequencies imply instability. According to equation \ref{letzte},
complex eigenfrequencies, and therefore instability, are obtained,
if ${\nabla}$ is finite and $n$ is sufficiently large. 
For fixed $n$ we obtain in the limit of large ${\nabla}$ (expansion of 
the root):
\begin{eqnarray}
  \sigma_1^2&=&\frac{1}{4}\frac{\pi^2n^2}{\psi^2(a-b)^2}
  \frac{1}{\left(\alpha-\frac{\delta\kappa_p}{(4-\kappa_T)}\right)}\label{root_1}\\
  \sigma_2^2&=&\tilde{\nabla}
  \left(\alpha-\frac{\delta\kappa_p}{(4-\kappa_T)}
  \right)
  -\frac{2\pi^2n^2}{\psi^2(a-b)^2}
  \frac{1}{\left(\alpha-\frac{\delta\kappa_p}{(4-\kappa_T)}
      \right)}\label{root_2}
\end{eqnarray}
Equation \ref{root_1} describes the eigenfrequencies of the decoupled shock modes 
discussed in section \ref{threezonemodel}, i.e., the second order analysis of 
the previous section is contained in the limit $\nabla\to \infty$
of the present approach. 
Instabilities described by equation \ref{letzte} resemble those of the averaged model,
rather than those given by equation \ref{root_1} for positive values of $\kappa_p$.
We conclude that a finite but large value of the stratification
parameter $\nabla=\frac{{\rm{d}}\log T}{{\rm{d}}\log p}$ is essential for  
instability. 
However, assuming $\nabla \to \infty$, which was done in the investigation 
of the three-zone-model (section \ref{threezonemodel}), is an oversimplification.

\section{CONCLUSIONS}

When following the non linear evolution of strange mode instabilities in the envelopes of 
massive stars, shock fronts were observed to be captured in the H-ionisation zone
some pulsation periods after reaching the non linear regime.
This effect is not observed in models of very hot envelopes (such as the massive star model 
investigated by \citet{GKCF99}), due to hydrogen
being ionised completely. The shocks trapped in the H-ionisation zone perform high
frequency oscillations (associated with the sound travel times across the 
shock zone) confined to its very vicinity, whereas the remaining parts of
the envelope vary on the dynamical timescale of the primary, strange mode instability.
By performing an appropriate linear stability analysis the high frequency 
oscillations were shown to be due to a physical instability, rather than being 
a numerical artifact.

An analytical model for the secondary, shock zone instabilities has been constructed.
As a result, high values of $\nabla$ were found to be responsible for instability. 
Contrary to the common stratification (convective, Rayleigh-Taylor) instabilities
driven by buoyancy forces and thus associated with (non radial) gravity modes, however,
the instabilities found here are associated with spherically symmetric acoustic waves.
An extension of the stability analysis to non radial perturbations would be instructive,
since we expect the acoustic instabilities identified here - similar to strange mode
instabilities (see \citet{GM96}) - 
not to be restricted to spherical geometry. Such an investigation 
would also reveal buoyancy driven instabilities, which we believe not to be relevant
for the following reasons: Their typical 
timescale is much longer than that of the acoustic instabilities, which will therefore
dominate the dynamics. Moreover, in addition to gravity, the acceleration 
due to the shock's velocity field
has to be taken into account and is likely to stabilize the stratification with respect
to convective instabilities. With respect to the aim of this paper to identify
the secondary, shock oscillations and their origin, a non radial analysis is 
beyond the scope of the present investigation and will be the subject of a forthcoming
publication. 

Since the oscillations are a physical phenomenon - rather than a numerical artifact -
they should not be damped by increasing the artificial viscosity as one would 
neglect a physical process whose influence on the long term behaviour of the system 
cannot yet be predicted. On the other hand, following the shock oscillations
by numerical simulation for more than a few dynamical timescales 
is not feasible due to the small timesteps necessary to 
resolve them. The confinement of the oscillations to the very vicinity of hydrogen 
ionisation, however, indicates a solution of the problem by means of domain decomposition:
The stellar envelope is decomposed into three domains: below, around and above the
shock. Only the narrow shock region needs high time resolution, the inner and outer
zones merely require the dynamical timescale to be resolved. The development of a 
code following this strategy is in progress.

Even if the appearance of the shock oscillations has so far prevented us from 
performing simulations in excess of several dynamical timescales, the velocity
amplitudes reach a significant fraction of the escape velocity. This indicates 
that pulsationally driven mass loss may be found in appropriate simulations.
Whether the new code will allow for the corresponding  long term 
simulations and thus possibly for the determination of mass loss rates, remains to be seen. 
Preliminary results will be published in a 
forthcoming paper. 

\section*{Acknowledgments} 

We thank Professor K.J. Fricke for encouragement and support.
Financial support by the Graduiertenkolleg 
``Str\"omungsinstabilit\"aten und Turbulenz'' (MG) and 
by the DFG under grant WA 633 12-1 (SC) is gratefully acknowledged. The
numerical computations have been carried out using the facilities  
of the GWDG at G\"ottingen.

\bsp

\label{lastpage}

\end{document}